\documentclass[final,3p,times]{elsarticle}

\usepackage{amsmath,amssymb,amsfonts,amscd,amsthm}
\usepackage{graphicx}

\usepackage{color,soul}

\newcommand{\R}{\mathbb{R}}
\newcommand{\D}{\mathrm{d}}
\newcommand{\e}{\mathrm{e}}
\newcommand{\JJ}{\mathcal{J}}
\newcommand{\pf}{\noindent{\sl Proof: }}
\newcommand{\QED}{$\Box$}
\newtheorem{claim}{Claim}[section]
\newtheorem{theorem}[claim]{Theorem}
\newtheorem{proposition}[claim]{Proposition}
\newtheorem{corrolary}[claim]{Corrolary}
\newtheorem{remark}[claim]{Remark}

\begin{document}

\begin{frontmatter}

\title{On the ground state of quantum graphs with attractive $\delta$-coupling}
\author[label1,label2]
{Pavel Exner\corref{cor1}}
\ead{exner@ujf.cas.cz}
\author[label1,label3]
{Michal Jex}
\ead{michal.j@centrum.cz}
\address[label1]{Doppler Institute for Mathematical Physics and Applied
Mathematics, Czech Technical University,
B\v rehov{\'a} 7, 11519 Prague, Czechia}
\address[label2]{Department of Theoretical Physics, Nuclear Physics
Institute, Czech Academy of Sciences,
25068 \v{R}e\v{z} near Prague, Czechia; phone +420-266-173-293, fax +420-220-940-165}
\address[label3]{Department of Physics, Faculty of Nuclear Sciences and Physical Engineering, Czech Technical University, B\v{r}ehov\'{a} 7, 11519 Prague, Czechia}
\cortext[cor1]{corresponding author}

\date{\today}
\begin{abstract}
We study relations between the ground-state energy of a quantum graph Hamiltonian with attractive $\delta$ coupling at the vertices and the graph geometry. We derive a necessary and sufficient condition under which the energy increases with the increase of graph edge lengths. We show that this is always the case if the graph has no branchings while both energy increase and decrease are possible for graphs with a more complicated topology.
\end{abstract}

\end{frontmatter}


\section{Introduction}

Quantum graphs proved themselves to be a class of systems offering numerous problems interesting from both the physical and mathematical point of view; we refer to the proceedings volume \cite{EKST} for an extensive bibliography. In this Letter we address the question about relations between the ground-state energy of such a Hamiltonian and geometric properties of the underlying graph, in particular, the lengths of its edges.

A motivation to study this kind of problem is twofold. On the physics side it is, of course, the importance of the ground state as the one to which the system tends to relax when it loses energy due to an interaction with the environment. Since quantum graphs model various real physical systems it is natural to ask about the geometric configurations which are energetically the most favourable. At the same time, mathematically the problem represents a natural extension of the usual spectral-geometry studies of the relations between spectral properties of differential operators and geometry of the manifolds supporting them.

We restrict here our attention to graphs with a finite number of edges, some of which may be semi-infinite, and an attractive $\delta$ coupling at the vertices, assuming that the motion at the graph edges away from the vertices is free. Such systems have always a nontrivial negative spectrum with a well-defined ground state; we will ask how the corresponding eigenvalue depends on the finite-edge lengths. First we analyze the case of $n$ attractive $\delta$ interactions on the line which can be regarded as a simple chain graph. We will prove that the ground-state energy moves up with increasing distances between the $\delta$ potentials in two different ways, by means of a Neumann bracketing and by using the well-known explicit form of such a Hamiltonian resolvent.

After that we will pass to general quantum graphs of the described class. We will show that in such a case the dependence on the edge length is more complicated and its sign is uniquely determined by the form of the ground-state eigenfunction on the particular edge. As long as the graph is a chain we have the monotonicity described above. On the other hand, we will give an example showing that once the graph has at least one nontrivial branching, i.e. a vertex of degree exceeding two, it is possible that the ground-state energy decreases with the increasing edge lengths.

Before proceeding further let us note that relations between quantum graph eigenvalues and edge lengths have been discussed also in other contexts. In particular, Friedlander \cite{Fr} derived a lower bound on higher eigenvalues for finite graphs in terms of the total graph size. On the other hand, Berkolaiko and Kuchment \cite{BK} studied general relations between the point spectrum and the set of edge lengths and coupling constants.

\setcounter{equation}{0}
\section{A warm-up: $\delta$ interactions on a line}

Consider first a particle on a line with a finite number of $\delta$-interactions the Hamiltonian of which can be formally written as $-\frac{\D^2}{\D x^2} + \sum_{j=1}^n \alpha_j\delta(x-y_j)$. Following \cite{AGHH} we denote this operator as $-\Delta_{\alpha,Y}$ where $\alpha:=\{\alpha_1,\dots,\alpha_n\}$ and $Y:=\{y_1,\dots,y_n\}$.
We suppose that all the points $y_j$ are mutually distinct and the interactions are attractive, \mbox{$\alpha_j<0$}, $j=1,\dots,n$. Under this assumption the continuous spectrum of $-\Delta_{\alpha,Y}$ covers the positive halfline and the discrete spectrum in the negative part of the axis is non-empty, in particular, there is a ground-state eigenvalue $\lambda_0<0$ with a strictly positive eigenfunction $\psi_0$\footnote{See \cite[Thm.~II.2.1.3]{AGHH}, and also Theorem~\ref{simple gs} below.}; we ask how does $\lambda_0$ depend on the geometry of the set $Y$.

One can conjecture that the ground-state energy decreases when the point interactions are closer to each other. First we prove this claim under an additional assumption.

\begin{proposition} \label{th: linebrac}
Consider sets $Y_1,\,Y_2$ of the same cardinality such that $y_{j,1}<
y_{j,2}<\ldots< y_{j,n}\,,\: j=1,2$. Let there be an $i$ such that $y_{2,l} = y_{1,l}$ for $l=1,\ldots,i$ and $y_{2,l} = y_{1,l}+\eta$ for $l=i+1,\ldots,n$. Suppose further the ground-state eigenfunction of the $-\Delta_{\alpha,Y_1}$ satisfies $\psi'_0(y_{1,i}+)<0$ and $\psi'_0(y_{1,i+1}-)>0$. Then we have $\min
\sigma({-\Delta_{\alpha,Y_1}})\leq \min \sigma({-\Delta_{\alpha,Y_2}})$ for any $\eta>0$.
\end{proposition}

\pf Since $\psi_0$ is positive and satisfies $\psi_0''= -\lambda_0 \psi_0$ between the point interaction sites, the function is convex; by the assumption there is then a point $x_0\in(y_{1,i},y_{1,i+1})$ such that $\psi'_0(x_0)=0$. Consider now the operator $-\tilde\Delta_{\alpha,Y_1}$ which acts as $-\Delta_{\alpha,Y_1}$ with the additional splitting\footnote{Adding a Neumann condition is understood here in the way standard in bracketing arguments \cite[Sec.~XIII.15]{RS}. Nevertheless, since Neumann condition is sometimes used as a synonym for Kirchhoff coupling in quantum graphs, we say ``splitting'' to stress that the functions from the domain of $-\tilde\Delta_{\alpha,Y_1}$ are in general discontinuous at $x_0$.} Neumann condition at the point $x_0$; it is obvious that the two operators have the same ground state. Such a Neumann condition separates the two halflines, hence  $-\tilde\Delta_{\alpha,Y_1}$ can be written as $-\tilde\Delta_{\alpha,Y_1}^l \oplus -\tilde\Delta_{\alpha,Y_1}^r$. Consider now the operator $-\hat\Delta_{\alpha,Y_2} := -\tilde\Delta_{\alpha,Y_1}^l \oplus -\Delta_N \oplus -\tilde\Delta_{\alpha,Y_1}^r$ where the added operator is the Neumann Laplacian on $L^2(0,\eta)$; it is clear that the latter does not contribute to the negative spectrum, hence $\min\sigma(-\hat\Delta_{\alpha,Y_2}) = \min\sigma( -\tilde\Delta_{\alpha,Y_1})$. Furthermore, $-\hat\Delta_{\alpha,Y_2}$ is obviously unitarily equivalent to $-\Delta_{\alpha,Y_2}$ with added splitting Neumann conditions at the points $x=x_0, x_0+\eta$, hence the sought result follows from Neumann bracketing \cite[Sec.~XIII.15]{RS}. \quad \QED

\medskip

It is not difficult to see that the assumption about the derivative signs is satisfied if $-\alpha_i,-\alpha_{i+1}$ are large enough or, which is the same by scaling, the distance $y_{i+1}-y_i$ is large enough. However, we can make a stronger claim without imposing restrictions on the ground-state eigenfunction derivatives.

\begin{theorem} \label{th: genline}
Suppose again that $\#Y_{1}=\#Y_{2}$ and $\alpha_j<0$ for all $j$. Let further $y_{1,i}-y_{1,j}\leq y_{2,i}-y_{2,j}$ hold for all $i,j$ and $y_{1,i}-y_{1,j}< y_{2,i}-y_{2,j}$ for at least one pair
of $\,i,j$, then we have $\min\sigma({-\Delta_{\alpha,Y_{1}}})< \min
\sigma({-\Delta_{\alpha,Y_{2}}})$.
\end{theorem}

\pf We employ Krein's formula \cite[Sec.II.2.1]{AGHH} which makes it possible to reduce the spectral problem at energy $k^2$ to solution of the secular equation, $\det\Gamma_{\alpha,Y}(k)=0$, where
 $$
[\Gamma_{\alpha,Y}(k)]_{jj'}=-[\alpha_{j}^{-1}\delta_{jj'} +G_{k}(y_{j}-y_{j'})]^{N}_{j,j'=1}
 $$
and $G_{k}(y_{j}-y_{j'})=\frac{i}{2k}\e^{ik\left|y_{j}-y_{j'}\right|}$ is the free resolvent kernel. Writing conventionally $k=i\kappa$ with $\kappa>0$, we have to investigate the \emph{lowest} eigenvalue of $\Gamma_{\alpha,Y}(i\kappa)$ which is, of course, given by
 $$
   \mu_0(\alpha,Y;\kappa) = \min_{|c|=1}\,  \left(c, \Gamma_{\alpha,Y}(i\kappa)c \right)
 $$
with the minimum taken over all $c\in\mathbb{C}^n$ with $|c|=1$. It is easy to see that $\mu_0(\alpha,Y;\kappa)>0$ for all $\kappa$ large enough; the ground state energy $-\kappa^2$ corresponds to the \emph{highest} value of $\kappa$ such that $\mu_0(\alpha,Y;\kappa)=0$. Since $[\Gamma_{\alpha,Y}(i\kappa)]_{ij}= -\delta_{ij}\alpha_i^{-1} - \frac{1}{2\kappa}\, \mathrm{e}^{-\kappa \ell_{ij}}$, where $\ell_{ij}=|y_i-y_j|$, the quantity to be minimized is explicitly
 $$
   \left(c, \Gamma_{\alpha,Y}(i\kappa)c \right) = \sum_{i=1}^n |c_i|^2 \left(-\frac{1}{\alpha_i}- \frac{1}{2\kappa} \right) - 2 \sum_{i=1}^n \sum_{j=1}^{i-1} \mathrm{Re\,} \bar c_i c_j \frac{\mathrm{e}^{-\kappa \ell_{ij}}}{2\kappa}
 $$
Next we notice that the eigenfunction corresponding to the ground state, i.e. $c$ for which the minimum is reached can be chosen \emph{strictly positive}; we write symbolically $c>0$ meaning $c_i>0,\: i=1,\dots,n$. This follows from the fact that the semigroup $\{ \mathrm{e}^{-t\Gamma_{\alpha,Y} (i\kappa)}:\: t\ge0 \}$ is positivity improving, as a consequence of strict negativity of the off-diagonal elements of $\Gamma_{\alpha,Y}(i\kappa)$ --- cf. \cite{RS}, Sec. XIII.12 and Problem XIII.97. This means, in particular, that we have
 $$
   \mu_0(\alpha,Y;\kappa) = \min_{|c|=1, c>0}\,  \left(c, \Gamma_{\alpha,Y}(i\kappa)c \right)
 $$
Take now two configurations, $(\alpha,Y)$ and $(\alpha,\tilde Y)$ such that $\ell_{ij} \le \tilde \ell_{ij}$ and the inequality is strict for at least one pair $(i,j)$. For any fixed $c>0$ we then have $\left(c, \Gamma_{\alpha,Y}(i\kappa)c \right) < \left(c, \Gamma_{\alpha,\tilde Y}(i\kappa)c \right)$, and consequently, taking a minimum overs all such $c$'s we get
 $$
   \mu_0(\alpha,Y;\kappa) < \mu_0(\alpha,\tilde Y;\kappa)
 $$
for \emph{any} $\kappa>0$ with the obvious consequence for the ground state of $-\Delta_{\alpha,Y}$; the sharp inequality in the last formula holds due to the fact that there is a $c$ for which the minimum is attained. \quad \QED

\begin{remark} \label{loop}
{\rm The argument used above can be extended to other situation. Take for instance, point interactions on a loop, in other words, on a finite interval with periodic boundary conditions. The corresponding Green's function is
 $$
 G_{i\kappa}(x,y) = \frac{\cosh \kappa(\ell-|x-y|)}{2\kappa \sinh \kappa\ell}\,, \quad |x-y|\le \frac12\ell\,,
 $$
where $\ell$ is the length of the loop. Writing the corresponding secular equation we find that expanding the loop without reducing the distances between the neighbouring point interaction sites means moving the ground-state energy up.
}
\end{remark}

\setcounter{equation}{0}
\section{Quantum graphs: setting the problem}

After this preliminary let us pass to a more general situation when the particle lives on a graph and the attractive point interaction represent couplings at the graph vertices. Consider a graph $\Gamma$ consisting of a set of vertices
$\mathcal{V}=\{\mathcal{X}_j: j\in I\}$, a set of finite edges
$\mathcal{L} =\{\mathcal{L}_{jn}: (j,n) \in I_\mathcal{L} \subset I\times I\}$ where $\mathcal{L}_{jn}$ is the the edge\footnote{Without loss of generality we may suppose that each pair of vertices is connected by a single edge; in the opposite case we add extra vertices of degree two to the ``superfluous'' edges and impose Kirchhoff conditions there.} connecting the vertices $\mathcal{X}_j$ and $\mathcal{X}_n$, and a set of infinite edges ${\mathcal L_\infty} =
\{\mathcal{L}_{k\infty}: k \in I_\mathcal{C}\}$ attached to them. We regard it as a configuration space of a quantum system with the Hilbert space
 $$
\mathcal{H} = \bigoplus_{j \in I_\mathcal{L}} L^2([0,l_{j}])\oplus
\bigoplus_{k \in I_\mathcal{C}} L^2([0,\infty))\,.
 $$
the elements of which can be written as columns $\psi  = \{(\psi_{jn}:
\mathcal{L}_{jn} \in \mathcal{L}\},\, \{\psi_{k\infty}: \mathcal{L}_{k\infty}\in
\mathcal{L}_\infty\})^T$. We consider the dynamics governed by a
Hamiltonian which acts as $-\mathrm{d}^2/\mathrm{d} x^2$ on each edge. In order to make it a self-adjoint operator, in general boundary conditions
 \begin{equation} \label{gen_bc}
   (U_j -I) \Psi_j +i (U_j +I) \Psi_j' = 0\,
 \end{equation}
with unitary matrices $U_j$ have to be imposed  at the vertices
$\mathcal{X}_j$, where $\Psi_j$ and $\Psi_j'$ are vectors of the
functional values and of the (outward) derivatives at the
particular vertex, respectively \cite{GG, Ha, KS}. In other words, the domain of the Hamiltonian consists of all functions in
$W^{2,2}(\mathcal{L}\oplus\mathcal{L}_\infty)$ which satisfy the conditions (\ref{gen_bc}). We will be interested in the following particular class:

\begin{itemize}

\item the internal part of the graphs is finite and so is the
number of external edges, $\# I_\mathcal{L}<\infty$ and $\#
I_\mathcal{C}<\infty$

\item  the coupling at each vertex is of $\delta$ type in terminology of \cite{E}, i.e. $U_j= {2\over n_j+i\alpha_j}\JJ-I$, where $n_j$ is the degree of the vertex $\mathcal{X}_j$ and $\JJ$ is the matrix having all the entries equal to one. Explicitly the coupling conditions \eqref{gen_bc} then become
 \begin{equation} \label{delta_bc}
   \psi_{j,i}(0)=\psi_{j,k}(0)=:\psi_j(0)\,, \quad j\in I,\,k=1,\dots,n_j\,, \quad\;
   \sum_{i=1}^{n_j} \psi_{j,i}'(0) = \alpha_j \psi_j(0)\,,
 \end{equation}
where each edge emanating from $\mathcal{X}_j$ is parametrized in such way that $x=0$ corresponds to the vertex

\item for ``free endpoints'', or vertices of degree one, parametrized by $x_j=l_j$, this in particular means the Robin condition, $\psi_j'(l_j) + \alpha_j \psi_j(l_j) = 0$

\item all the couplings involved are non-repulsive, $\alpha_j\le 0$ for all $j\in I$, and at least one of them is attractive, $\alpha_{j_0}<0$ for some $j_0\in I$

\end{itemize}

In such a case it is not difficult to express the quadratic form associated with the quantum-graph Hamiltonian $H$: it is given by\footnote{A meticulous reader might notice that numberings of the functions on the edges differ; sometimes it is practical to number the edges, sometimes vertices at their endpoints, or edges sprouting from a given one. We are sure that this can cause no misunderstandings.}
 \begin{equation} \label{qform}
 q[\Psi] = \sum_{j \in I_\mathcal{L}} \int_0^{l_j} |\psi'_j(x)|^2
   \,\D x + \sum_{k \in I_\mathcal{C}} \int_{\R_+} |\psi'_k(x)|^2
   \,\D x + \sum_{i \in I} \alpha_i |\psi_i(0)|^2 ,
 \end{equation}
where $\psi_j,\, \psi_k$ are components of the wave function
$\Psi$  on the internal and external edges, respectively, and
$\psi_i(0)$ are the values at the vertices. The domain of the form consists of $L^2$ functions which are $W^{1,2}$ on the graph edges and continuous at the vertices.

\begin{proposition}
$\inf\sigma(H)<0$ holds under the stated assumptions.
\end{proposition}

\pf If $I_\mathcal{C}= \emptyset$ we take a constant function,
$\Psi=c$ on $\Gamma$ which belongs to the form domain because
$\Gamma$ has then a finite length; we get $q[\Psi]\le \alpha_{j_0} |c|^2$. On the other hand, if $I_\mathcal{C}\ne\emptyset$, we take $\Psi$ equal to $c$ on the internal part of the graph and to $\psi_k(x) = c\,\e^{-\kappa x}$ on each external semi-infinite edge. The integrals over the internal edges vanish as before and those over external ones are easily evaluated; we get
 $$
q[\Psi]\le \left( \alpha_{j_0} +\frac12 \kappa \#I_\mathcal{C}
\right) |c|^2
 $$
which can be made negative by choosing $\kappa$ small enough.
\quad \QED

\begin{theorem} \label{simple gs}
In addition, let $\Gamma$ be connected, then the bottom of the spectrum $\lambda_0=\inf\sigma(H)$ is a simple isolated eigenvalue. The corresponding eigenfunction $\Psi^{(0)}$ can be chosen strictly positive on $\Gamma$ being convex on each edge.
\end{theorem}

\pf Consider a disjoint graph with all the vertex couplings changed to Dirichlet conditions. In such a case the spectrum is positive; it is discrete if $I_\mathcal{C}=\emptyset$ and equal to $\R_+$ otherwise. By Krein's formula \cite[Proposition~2.3]{P}, the original operator differs from the Dirichlet decoupled one by a finite-rank perturbation in the resolvent, hence their essential spectra are the same by Weyl's theorem and the negative spectrum of $H$ may consist at most of a finite number of eigenvalues of finite multiplicity; by the previous proposition it is nonempty and the ground-state eigenvalue exists.

The ground state positivity follows, e.g., from a quantum-graph modification of the Courant theorem \cite{BBRS}. The eigenfunction being positive and its component $\psi^{(0)}_j$ at the $j$th edge twice differentiable away of the vertices, we have $(\psi^{(0)}_j)'' = -\lambda_0 \psi^{(0)}_j>0$, which means the convexity. \quad \QED

\medskip

In fact, one can say more about the ground-state eigenfunction because the corresponding Schr\"odinger equation can be solved explicitly. Writing the spectral threshold as $\lambda_0= -\kappa^2$ we see that the eigenfunction component on each edge is a linear combination of $\e^{\kappa x}$ and $\e^{-\kappa x}$. Since we are free to choose the edge orientation, each component has one of the following three forms,
 \begin{equation} \label{component_form}
 \psi^{(0)}_j(x) = \left\{
 \begin{array}{ll} c_j\cosh\kappa(x+d_j)\,, &\qquad d_j\in\R \\
 c_j\,e^{\pm\kappa(x+d_j)}\,, &\qquad d_j\in\R \\
 c_j\sinh\kappa(x+d_j)\,, &\qquad x+d_j> 0 \end{array} \right.
 \end{equation}
where $c_j$ is a positive constant. For further purposes we introduce \emph{edge index}
 \begin{equation} \label{edge_index}
 \sigma_j:= \left\{
 \begin{array}{rll} +1 &\quad\dots\quad& \psi^{(0)}_j(x)=c_j\cosh\kappa(x+d_j)\\
 0 &\quad\dots\quad& \psi^{(0)}_j(x)=c_j\,e^{\pm\kappa(x+d_j)} \\
 -1 &\quad\dots\quad& \psi^{(0)}_j(x)=c_j\sinh\kappa(x+d_j) \end{array} \right.
 \end{equation}

\setcounter{equation}{0}
\section{Monotonicity proof by a scaling argument}

From now on we consider connected graphs only which we can do without loss of generality, since otherwise we deal with each connected component separately. By Theorem~\ref{simple gs} the graph's Hamiltonian $H$ then has a simple ground state with positive eigenfunction. Using the above definition, we can compare graphs with the same index structure.

Given $\Gamma$ and $\tilde\Gamma$ with the same topology differing possibly by inner edge lengths, we consider the family of interpolating graphs having the length of the $j$th edge in the closed interval with $l_j$ and $\tilde l_j$ as endpoints\footnote{This family corresponds to a closed parallelopiped in the natural parameter space $(0,\infty)^{\sharp I_\mathcal{L}}$ of our problem given by the interior edge lengths. If at least one of the edges has the same length in both the $\Gamma$ and $\tilde\Gamma$ the said parallelopiped is degenerate and we regard it instead as a subset in the reduced parameter space referring to the changing edge lenghts. The eigenvalues are analytic functions on the interior of such a set -- cf.~\cite{BK}.}. We say that the graphs $\Gamma$ and $\tilde\Gamma$ belongs to the same \emph{ground-state class} if the edge indices of the graphs edges remain the same for this whole family. Equipped with this notion we can make the following claim.

\begin{theorem} \label{scaling arg}
Under the stated assumptions, consider graphs $\Gamma$ and $\tilde\Gamma$ of the same ground-state class. Let $H$ and $\tilde H$ be the corresponding Hamiltonians with the same couplings in the respective vertices, and $\lambda_0$ and $\tilde\lambda_0$ the corresponding ground-state eigenvalues. Suppose that $\sigma_j\tilde l_j\le \sigma_j l_j$ holds all $j\in I_\mathcal{L}$ such that $|\sigma_j|=1$ and $\tilde l_j=l_j$ if $\sigma_j=0$, then $\tilde\lambda_0 \le \lambda_0$; the inequality is sharp if $\sigma_j\tilde l_j< \sigma_j l_j$ holds for at least one $j\in I_\mathcal{L}$.
\end{theorem}

\pf It is obviously sufficient to compare graphs differing just by the length of a single inner edge corresponding to a fixed index value $j\in I_\mathcal{C}$ with $|\sigma_j|=1$, and it is enough to prove the claim locally. We choose a finite-length segment $J\equiv[a,b]$ in the interior of the $j$th edge and write $\Gamma$ as the union of $J$ and $\Gamma_J:= \Gamma\setminus J$. Without loss of generality we may choose $J$ in such a way that $b-a> l_j-\tilde l_j$ if $\sigma_j=1$ and $b-a<l_j$ if $\sigma_j=-1$. Then $\tilde\Gamma$ can be written as $\Gamma_J\cup \tilde J$ where $\tilde J$ is obtained by scaling of $J$ with the factor $\xi:=|\tilde J||J|^{-1}$ being less than one in case of a shrinking edge and larger than one otherwise. In order to prove the desired result we have to find a function $\Psi\in L^2(\tilde\Gamma)$ such that the Rayleigh quotient on the tilded graph satisfies
 \begin{equation} \label{rayleigh}
 \frac{\tilde q[\Psi]}{\|\Psi\|^2} < \lambda_0\,.
 \end{equation}
We construct such a trial function $\tilde\Psi^{(0)}$ in the following way: we put $\tilde\Psi^{(0)}(x) = \Psi^{(0)}(x)$ for $x\in \Gamma_J$ while the $j$th component on $\tilde J$ is obtained by scaling
 \begin{equation} \label{trialscale}
 \tilde\psi_j^{(0)}(\tilde a+\xi y) = \psi_j^{(0)}(a+y) \quad \mathrm{for} \quad 0 \le y \le |J|\,;
 \end{equation}
in order to prove (\ref{rayleigh}) we have to choose $\xi<1$ \emph{iff} $\sigma_j=1$ and vice versa. The Rayleigh quotient for the function (\ref{trialscale}) can be easily rewritten in a natural notation as
 \begin{equation} \label{rayleigh2}
 \frac{\tilde q[\tilde\Psi^{(0)}]}{\|\tilde\Psi^{(0)}\|^2} = \frac{a+b\xi^{-1}}{c+d\xi} =: f(\xi)\,,
 \end{equation}
where
 $$
 a:= q_{\Gamma_J}[\Psi^{(0)}]\,, \quad b:= \int_J |(\psi^{(0)}_j)'(x)|^2\, \mathrm{d}x\,,
 $$
and $c,d$ are the parts of the squared norm of $\Psi^{(0)}$ corresponding to $\Gamma_J$ and $J$, respectively. It is enough to check that $\sigma_j f'(1)= -\sigma_j(bc+2bd+ad)(c+d)^{-2}>0$. Choosing the ground-state eigenfunction $\Psi^{(0)}$ conventionally with the norm equal to one, we have $c+d=1$ and $a+b=\lambda_0$, hence the property to be checked is $-\sigma_j(\lambda_0 d+b)>0$, or more explicitly
 $$
 -\sigma_j\left(\lambda_0 \|\psi_j^{(0)}\|^2_J + \|(\psi_j^{(0)})'\|^2_J \right) > 0\,.
 $$
Using $\lambda_0= -\kappa^2$ we find for $\sigma_j=1$
 \begin{eqnarray*}
 \lefteqn{\int_J |(\psi^{(0)}_j)'(x)|^2\, \mathrm{d}x = c_j^2\kappa^2 \int_J (\sinh\kappa x)^2\, \mathrm{d}x < c_j^2\kappa^2 \int_J (\cosh\kappa x)^2\, \mathrm{d}x} \\ && = -\lambda_0 \int_J |\psi^{(0)}_j(x)|^2\, \mathrm{d}x\,, \phantom{AAAAAAAAAAAAAAAAAAAAA}
 \end{eqnarray*}
and the opposite inequality for $\sigma_j=-1$ where the roles of hyperbolic sine and cosine are interchanged. Hence inequality (\ref{rayleigh}) is satisfied for for $\xi<1$ if $\sigma_j=1$ and $\xi>1$ if $\sigma_j=-1$, provided $|1-\xi|$ is small enough, which is what we have set out to prove. \quad \QED

\begin{remark} \label{sigma_zero}
{\rm The case $\sigma_j=0$ is nontrivial; the critical case in the example given in the next section shows that such a possibility is not excluded in case of finite edges and the ground-state energy can be independent of length changes of such edges, however, a more subtle analysis is needed to treat such situations in general.
}
\end{remark}

\setcounter{equation}{0}
\setcounter{figure}{0}
\section{Discussion}

Let us now ask what consequences can one derive from our main result given in Theorem~\ref{scaling arg}. First we notice that the graphs without branchings belong all to the same class and one is able to extend to them conclusions of Theorem~\ref{th: genline} and Remark~\ref{loop}. Specifically, we can make the following claim.

\begin{corrolary} \label{nobranch}
In the setting of Theorem~\ref{scaling arg} suppose that the graph $\Gamma$ has no branchings, i.e. the degree of no vertex exceeds two. Then the index of any edge is non-negative being equal to one for any internal edge. Consequently, a length increase of any internal edge moves the ground-state energy up.
\end{corrolary}

\pf By assumption a graph without branchings is a chain of edges, either closed into a loop or open; in view of Remark~\ref{loop} we can consider only the latter possibility.
It is obvious that it is not possible that all the edges have negative index. If the first or the last one are semi-infinite, their index must be zero; if all the edges are finite the attractive $\delta$-coupling would require that $\mathrm{sgn}\, \psi'_0$ remains unchanged over the whole chain and the non-repulsive Robin condition on one of the endpoints could not be satisfied. The question is whether one can have a $\sinh$-type solution at some position within the chain. In such a case there would be a vertex in which wavefunction components with different indices have to match. Let us parametrize the chain by a single variable $x$ choosing $x=0$ for the vertex in question. Suppose that the (non-normalized) ground-state eigenfunction equals $\psi_j(x)=\cosh\kappa(d_1-x)$ for $x<0$ and $\psi_{j+1}(x)=c\,\sinh\kappa(d_2\mp x)$ for $x>0$. By assumption they are coupled by an attractive $\delta$ interaction, hence $c$ is determined by the continuity requirement and $\psi'_{j+1}(0+)-\psi'_j(0-)$ must be negative; recall that the ground-state eigenfunction is positive. However, this expression equals $\mp\kappa \cosh\kappa(d_1\pm d_2)/ \sinh\kappa d_2$, hence the needed match is impossible for a $\sinh$ solution decreasing towards the vertex. The same is true for the opposite order of the two solutions, and  similarly one can check that a negative-index edge cannot neighbour with a semi-infinite one.  \quad \QED

\medskip

\begin{figure}[h]
\begin{picture}(0,0)
  \includegraphics[width=10cm, keepaspectratio]{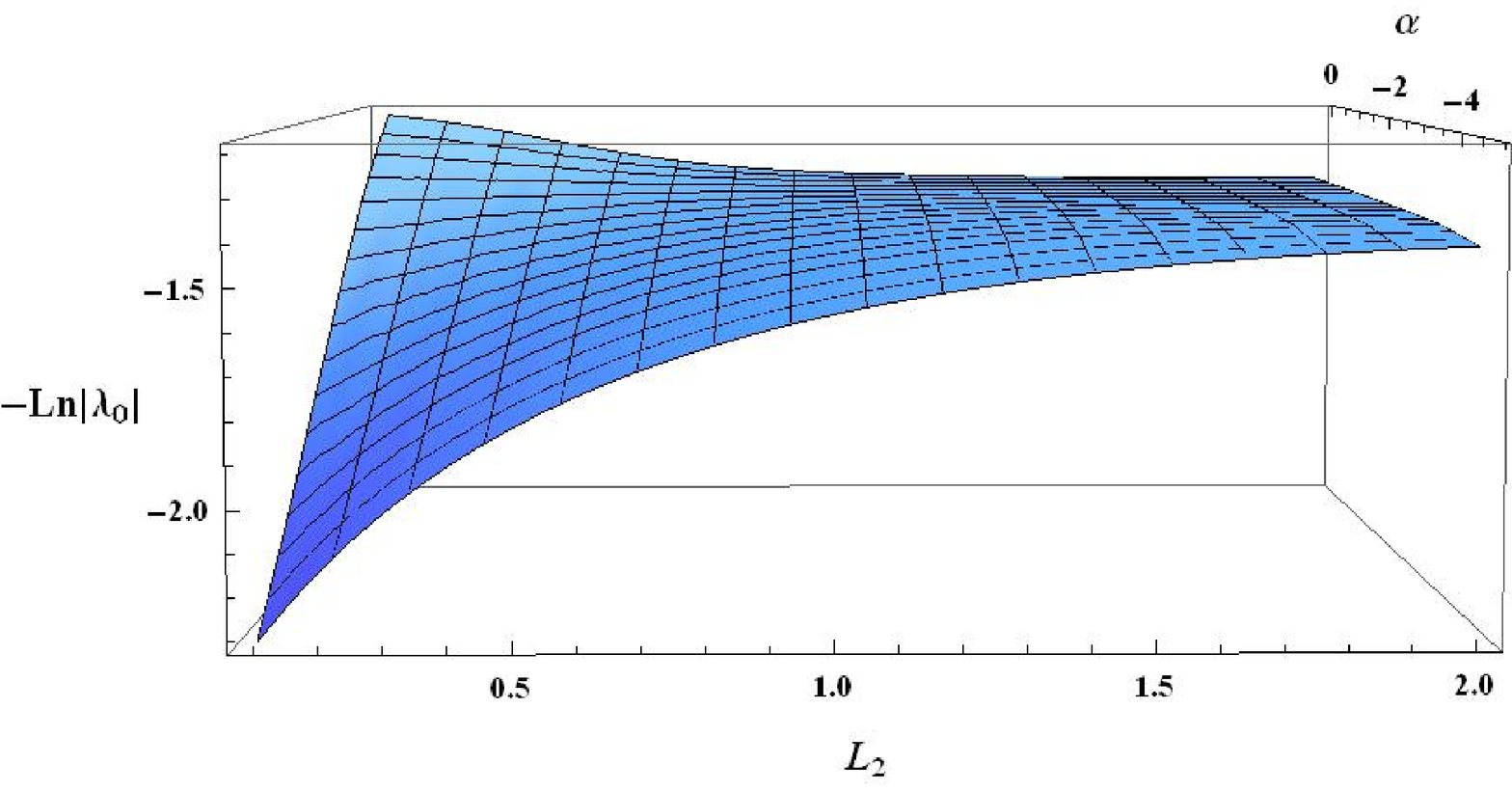}
\end{picture}%
\setlength{\unitlength}{4144sp}%
\begin{picture}(6431,2448)(79,-1732)

 \put(6216,-1096){\circle*{40}}
 \put(6216,-296){\circle*{40}}
 \put(6716,204){\circle*{40}}
 \put(5716,204){\circle*{40}}
 \thicklines
 \put(6216,-296){\line(0,-1){800}}
 \put(6216,-296){\line(1,1){500}}
 \put(6216,-296){\line(-1,1){500}}
 \put(6316,-346){\mbox{$\alpha$}}
 \put(5556,354){\mbox{$\alpha_1$}}
 \put(6456,354){\mbox{$\alpha_1=-1.5$}}
 \put(5966,-1296){\mbox{$\alpha_2=-2$}}
 \put(6266,-796){\mbox{$L_2$}}
 \put(5666,-106){\mbox{$L_1$}}
 \put(6616,-106){\mbox{$L_1=1$}}

\end{picture}%
\caption{The ground-state energy of sketched star graph as a function of $L_2$ and $\alpha$.} \label{stargraph}
\end{figure}

On the other hand, for graphs with a more complicated topology the analogous claim is no longer true. We will illustrate it on a simple example of a star graph with mirror symmetry sketched on Fig.~\ref{stargraph}. We have plotted here the ground-state energy --- in the logarithmic scale to make the effect more visible --- as a function of the edge length $L_2$ and the coupling constant $\alpha$ in the central vertex. We see two regimes here. For weak attractive coupling, $\alpha_\mathrm{crit} < \alpha <0$ where $\alpha_\mathrm{crit} \approx -1.09088$, the ground-state energy decreases with increasing $L_2$ while the opposite is true if $\alpha< \alpha_\mathrm{crit}$; at the critical value the energy is independent of $L_2$ and the solution on the ``axial'' edge is a pure exponential.

The reason why this happens in the example is obvious. The mirror symmetry allows us to decompose the problem into a symmetric part, where the ground state is to be sought, and antisymmetric one which reduced trivially to the Dirichlet problem on a single interval. Using the notation from the proof of Corollary~\ref{nobranch}, the left-hand side of the derivative condition in the symmetric part equals $\psi'_2(0+)-2\psi'_1(0-)$, hence the argument used there no longer applies. On the other hand, it is not difficult to construct examples without a symmetry in which we have different regimes; an open question is whether one can find a general regime characterization for an arbitrary branching graph.

Let us finally recall that our main result, Theorem~\ref{scaling arg}, holds for graphs belonging to the same ground-state class. We know from \cite{BK} that the notion is not empty: it follows from eigenfunction dependence on the edge lengths that a given graph has a neighbourhood in the parameter space where the indices do not change. The question about existence of different classes for graphs of the same topology is open and interesting.

\section*{Acknowledgments}
We are grateful to the referees for their comments which helped us to improve the text. The research was supported by the Czech Ministry of Education, Youth and Sports, and Czech Science Foundation within the projects LC06002 and P203/11/0701.



\begin{thebibliography}{article}

\setlength{\itemsep}{0pt}

\bibitem{EKST}
P.~Exner, J.~Keating, P.~Kuchment, T.~Sunada, and A.~Teplyaev: \emph{eds.}, ``Analysis on Graphs and its Applications'', Proc. Symp. Pure Math., (AMS Chelsea, Rhode Island, 2008).

\bibitem{Fr}
L. Friedlander: Extremal properties of eigenvalues for a metric graph, {\it Ann. Inst. Fourier} \textbf{55} (2005), 199--211.

\bibitem{BK}
G. Berkolaiko, P. Kuchment: Dependence of the spectrum of a quantum graph on vertex conditions and edge lengths, {\it AMS Contemp. Math.}, to appear; \texttt{arXiv:1008.0369v2}.

\bibitem{AGHH}
S.~Albeverio, F.~Gesztesy, R.~Hoegh-Krohn, and H.~Holden:
``Solvable Models in Quantum Mechanics'', 2nd ed. with appendix by P. Exner, (AMS Chelsea, Rhode Island, 2005).

\bibitem{RS}
M.~Reed, B.~Simon: ``Methods of Modern Mathematical Physics IV: Analysis of operators'', (Academic press, New York 1978).

\bibitem{GG}
V.I. Gorbachuk, M.L. Gorbachuk: ``Boundary Value Problems for
Operator Differential Equations'', (Kluwer, Dordrecht 1991).

\bibitem{Ha}
M. Harmer: Hermitian symplectic geometry and extension theory,
{\it J. Phys. A: Math. Gen.} \textbf{33} (2000), 9193--9203.

\bibitem{KS}
V.~Kostrykin, R.~Schrader: Kirchhoff's rule for quantum wires,
\emph{J. Phys. A: Math. Gen.} \textbf{32} (1999), 595--630.

\bibitem{E}
P.~Exner: Contact interactions on graph superlattices, \emph{J. Phys. A: Math. Gen.} \textbf{29} (1996), 87--102.

\bibitem{P}
K.~Pankrashkin: Sur l'analyse de mod\`eles math\'ematiques
issus de la m\'ecanique quantique, \emph{Habilitation}, Univerit\'{e} Paris-Sud, Orsay 2010.

\bibitem{BBRS}
R.~Band, G.~Berkolaiko, H.~Raz, U.~Smilansky: On the connection between the number of nodal domains on quantum graphs and the stability of graph partitions, \texttt{arXiv:1103.1423}

   \end{thebibliography}
\end{document}